\documentstyle[prl,aps,epsf]{revtex}
\begin{document}

\title{Three-Body Halo Fragmentation: Polarization effects}
\author{{\sc E.~Garrido, D.V.~Fedorov and A.S.~Jensen} \\
{\it Institute of Physics and Astronomy, \\
Aarhus University, DK-8000 Aarhus C, Denmark}}
\date{\today}

\maketitle

\begin{abstract}
Momentum distributions of particles from nuclear break-up of fast
three-body halos are calculated for polarized beam and outgoing 
fragments. These momentum distributions are combined in  
observable quantities which emphasize the two-body
correlations in the initial three-body structure as well as in the final
two-body system. Applying the formalism to $^{11}$Li we find a sensitive 
dependence on the $p$-wave content and on the position of the $p$-resonances 
in $^{10}$Li. Polarization experiments are then a new tool to study 
correlations in nuclear halos. 

\vspace{5mm}



\end{abstract}

\vspace{5mm}

\paragraph*{Introduction.}

The unusual halo nuclei appear under conditions realized
at the neutron dripline \cite{han95}. The structure
of these nuclei is dominated by a configuration where an ordinary
nucleus (core) is surrounded by weakly bound neutrons. For Borromean systems 
\cite{dima1,zhukov}, three particles without bound two-body subsystems, 
specific correlations must be present to provide the total binding. The
nature of these correlations is rather difficult to investigate
and the data have almost always been consistent with uncorrelated
motion of particles around the core. The mixture of $s$- and
$p$-states in the neutron-core subsystem is an obvious manifestation of
the correlations. However its existence has been rather difficult to
confirm experimentally.

The major source of detailed information is the measurements of
particle momentum distributions after fragmentation reactions of these
halo nuclei \cite{anne,zinser,nilsson,humbert}. In the typical
experimental setup two of the three constituent particles are detected
after the high energy halo fragmentation in anticoincidence with the
third particle, which has been instantly captured by the target. The
sudden approximation or the Serber mechanism, equivalent to
instantaneous cease of the interactions between the constituent
particles, has been proved to be adequate in data interpretation
\cite{zhukov,thompson}. Furthermore, it has been pointed out recently that
inclusion of the interaction between the two remaining particles
greatly improves the model \cite{zinser,korshe,edu}.  

A polarized beam interacting with a target is a tool sensitive to components
of the wave function with non-zero relative angular momentum.
In particular for $^{11}$Li, the resulting
particle momentum distribution should be sensitive to the amount of
$p$-state admixture in the neutron-core relative state. More details and
strong signals are expected when the polarization of the fragments is 
simultaneously measured.

The purpose of this letter is to investigate the effects of polarization
on halo nuclear reactions in view of extracting information about
the correlations in these nuclei. We shall first briefly sketch the
new method for computation of fragment momentum distributions arising
from polarized beam experiments. We shall investigate the effects of 
polarized beams and unpolarized fragments, as well as polarized beams 
and polarized fragments. 

\paragraph*{Method.}
In the sudden approximation the transition matrix is proportional to the
overlap between the initial and the final wave functions. This
approximation is suitable for neutron halo fragmentation, as
first suggested for the deuteron stripping by Serber \cite{serber} and
later used in many connections \cite{satchler}. The projectile wave function is
denoted by $\Psi^{JM}(\mbox{\bf x}, \mbox{\bf y})$, where $J$ and $M$
are the total spin and its projection. The coordinates {\bf x} and {\bf y} 
are the usual Jacobi coordinates \cite{dima1}, where {\bf x} is drawn between 
the two particles surviving after the fragmentation. 
The final state wave function is given by 
$ e^{i \mbox{\scriptsize\bf k}_y \cdot \mbox{\scriptsize\bf y} }
\chi_{s_y}^{\sigma_y}
\sum_{s_x \sigma_x} 
\langle s_1, s_2; \sigma_1, \sigma_2 | s_x \sigma_x \rangle 
w^{s_x \sigma_x}(\mbox{\bf k}_x, \mbox{\bf x} )$,
where $w^{s_x \sigma_x}(\mbox{\bf k}_x, \mbox{\bf x} )$
describes the final two-body system with spin $s_x$ and projection
$\sigma_x$. The quantum numbers $s_1$, $s_2$, $_y$, $\sigma_1$,
$\sigma_2$, and $\sigma_y$ are the spins and projections of the three
particles, and the plane wave describes the motion of the center of
mass of the two-body system. The plane waves \cite{zhukov} correctly
describe the shapes of the high energy momentum distributions in the
region of the first maximum. By squaring this overlap, and subsequently
summing up over the initial and final magnetic substates we get the
following expression for the cross section or momentum distribution
\begin{eqnarray}
\frac{d^6 \,\sigma}{d\,\mbox{\bf k}_x d\,\mbox{\bf k}_y} &\propto&
\sum_M W_{\mbox{\scriptsize init}}(M) 
\sum_{\sigma_1} W_{\mbox{\scriptsize fin}}(\sigma_1) 
\sum_{\sigma_2} W_{\mbox{\scriptsize fin}}(\sigma_2)
\sum_{s_x, s^\prime_x, \sigma_x}
   \langle s_1, s_2; \sigma_1, \sigma_2 | s_x \sigma_x \rangle
   \langle s_1, s_2; \sigma_1, \sigma_2 | s^\prime_x \sigma_x \rangle
   \nonumber \\ & &
\sum_{\sigma_y}
\langle e^{i \mbox{\scriptsize\bf k}_y \cdot \mbox{\scriptsize\bf y} }
        \chi_{s_y}^{\sigma_y} 
        w^{s_x \sigma_x}(\mbox{\bf k}_x, \mbox{\bf x} )
|
        \Psi^{JM}(\mbox{\bf x}, \mbox{\bf y}) \rangle^\ast
\langle e^{i \mbox{\scriptsize\bf k}_y \cdot \mbox{\scriptsize\bf y} }
        \chi_{s_y}^{\sigma_y}
        w^{s^\prime_x \sigma_x}(\mbox{\bf k}_x, \mbox{\bf x} )
|
        \Psi^{JM}(\mbox{\bf x}, \mbox{\bf y}) \rangle
\label{eq4}
\end{eqnarray}
where $W_{\mbox{\scriptsize init}}(M)$, 
$W_{\mbox{\scriptsize fin}}(\sigma_1)$, and 
$W_{\mbox{\scriptsize fin}}(\sigma_2)$ give the probability of
occupation of each magnetic substate in the initial and final states.
When the beam is not polarized and there is no detection of final spin
states these probabilities take the values 1/(2$J$+1), 1/(2$s_1$+1), and 
1/(2$s_2$+1), and we obtain the unpolarized expression.

From {\bf x} and {\bf y} we define the hyperspherical coordinates
$\rho=\sqrt{x^2+y^2}$, $\alpha=\arctan (x/y)$, and $\Omega_x$ and
$\Omega_y$ giving the directions of {\bf x} and {\bf y}. Analogously, in
momentum space we define $\kappa=\sqrt{k_x^2+k_y^2}$, 
$\alpha_\kappa=\arctan(k_x/k_y)$, $\Omega_{k_x}$, and $\Omega_{k_y}$. 

The three-body wave function is obtained by solving the
Faddeev equations in coordinate space \cite{dima2}. This function is then
expanded in terms of the complete set of hyperspherical harmonics 
$\mbox{\bf Y}_{\ell_x,\ell_y}^{K,L}(\alpha,\Omega_x,\Omega_y)$,
where $\ell_x$ and $\ell_y$ are the orbital angular
momenta associated to the Jacobi coordinates {\bf x} and {\bf y}, 
and $L$ is the coupling of both momenta. The quantum number $K$ is 
usually called the hypermoment. The function 
$\Psi^{JM}(\mbox{\bf x}, \mbox{\bf y})$ can then be expanded as
\begin{equation}
\Psi^{JM}(\mbox{\bf x}, \mbox{\bf y})=
\sum_{K \ell_x \ell_y s_x L S} F_{\ell_x \ell_y s_x}^{K L S}(\rho)
\left[
  \mbox{\bf Y}_{\ell_x,\ell_y}^{K,L}(\alpha,\Omega_x,\Omega_y)
\otimes \chi^S_{s_x,s_y} \right]^{J M}
\label{expan}
\end{equation}
where $\chi^S_{s_x,s_y}$ is the spin function obtained 
by coupling the two-body spin $s_x$ and the spin $s_y$ of the third particle 
to the total spin $S$. 

We expand $w^{s_x \sigma_x}(\mbox{\bf k}_x, \mbox{\bf x} )$  
in partial waves
\begin{equation}
w^{s_x \sigma_x}(\mbox{\bf k}_x, \mbox{\bf x}) =
\sqrt{\frac{2}{\pi}} \frac{1}{k_x x} \sum_{j_x, \ell_x, m_x}
u_{\ell_x s_x}^{j_x}(k_x,x) {\cal Y}_{j_x \ell_x s_x}
                              ^{m_x^{\mbox{\normalsize $\ast$}}} (\Omega_x)
\sum_{m_{\ell_x}} 
      \langle \ell_x s_x ; m_{\ell_x} \sigma_x|j_x m_x\rangle
 i^\ell_x Y_{\ell_x m_{\ell_x}}(\Omega_{k_x}) ,
\label{eq2}
\end{equation}
where the radial functions $u_{\ell_x s_x}^{j_x}(k_x,x)$ are obtained 
by solving the Schr\"{o}dinger equation with the corresponding two-body 
potential. The quantum numbers contained in eq.(\ref{eq2}) correspond 
to the ones in the two-body subsystem in eq.(\ref{expan}).
Following this procedure we incorporate Final State Interactions (FSI)
into the calculation. When FSI is not considered the
function $w^{s_x \sigma_x}(\mbox{\bf k}_x, \mbox{\bf x})$ reduces to a plane
wave, and the overlaps in eq.(\ref{eq4}) become the Fourier transform of
the three-body wave function.

First we consider the case of 100\% polarized beam 
($W_{\mbox{\scriptsize init}}(M^\prime)=\delta_{M^\prime M}$). 
After substituting 
eqs.(\ref{expan}) and
(\ref{eq2}) into eq.(\ref{eq4}), and integrating analytically over
$\Omega_{k_y}$ and $\varphi_{k_x}$ ($\Omega_{k_x} \equiv (\theta_{k_x},
\varphi_{k_x}$)), we get the expression 
\begin{eqnarray}
\frac{d^3 \,\sigma}{d\,k_y d\,k^\perp_x d\,k^\parallel_x} & \propto &
 k_y^2 k_x^\perp \frac{2}{\pi^2}
\sum_I (-1)^{J-M} \left(
    \begin{array}{ccc}
       J & J & I \\
       M &-M & 0
    \end{array}
    \right) \hat{I}^2 P_I(\cos{\theta_{k_x}})
\sum_{j_x \ell_x s_x} \sum_{j_x^\prime j_y \ell_y}
\sum_{L S L^\prime S^\prime} 
                 \nonumber \\ & &
(-1)^{2s_y+J-s_x+j_x-j_x^\prime-j_y} 
{\cal I}_{\ell_x s_x j_x}^{\ell_y L S}(\kappa, \alpha_\kappa)
{\cal I}_{\ell_x s_x j_x^\prime}^{\ell_y L^\prime S^\prime}
                                                    (\kappa, \alpha_\kappa)
\hat{j}_x^2 \hat{j^\prime}_x^2 \hat{j}_y^2
\hat{\ell}^2_x \hat{J}^2 \hat{L} \hat{S}
\hat{L^\prime} \hat{S^\prime}   \nonumber \\ &&
\left(
    \begin{array}{ccc}
       \ell_x & \ell_x & I \\
       0 & 0 & 0
    \end{array}
                           \right)
\left\{
    \begin{array}{ccc}
       j_x & j^\prime_x & I \\
       J & J & j_y
    \end{array}
                           \right\}
\left\{
    \begin{array}{ccc}
       j_x & j^\prime_x & I \\
       \ell_x & \ell_x & s_x
    \end{array}
                           \right\}
\left\{
    \begin{array}{ccc}
       J & j_x & j_y \\
       L & \ell_x & \ell_y \\
       S & s_x & s_y
    \end{array}
                           \right\}
\left\{
    \begin{array}{ccc}
       J & j^\prime_x & j_y \\
       L^\prime & \ell_x & \ell_y \\
       S^\prime & s_x & s_y
    \end{array}
                           \right\}
\label{eq5}
\end{eqnarray}
where $k_y= |\mbox{\bf k}_y|$, $k^\perp_x=|\mbox{\bf k}_x| \sin{\theta_{k_x}}$, 
$k^\parallel_x=|\mbox{\bf k}_x| \cos{\theta_{k_x}}$, and
$\hat{a}=\sqrt{2a+1}$. The functions ${\cal I}_{\ell_x s_x j_x}^{\ell_y L
S}(\kappa, \alpha_\kappa)$ are only available as numerical functions. 
By integration over $k_y$ and either $k^\perp_x$ or $k^\parallel_x$, 
we obtain respectively 
${\cal Q}_1^M(k^\parallel_x)$ or ${\cal Q}_2^M(k^\perp_x)$, the 
one-dimensional or two-dimensional momentum distribution 
(note that in our description there is no preferred axis, and therefore the 
one-dimensional longitudinal and transverse momentum distributions are 
identical). 

From the definitions of 
the coordinates it is clear that ${\cal Q}_{1,2}^M$ expresses the relative
momentum distribution of the two particles in the final state. To
compare with the experimental data one has to refer the momentum
distributions to the center of mass of the three-body system. The
appropriate expressions, used in the numerical calculations, are rather 
complicated but the qualitative features remain unchanged. Details about
these expressions will be given in a forthcoming paper. 

 The summation over $M$ in eq.(\ref{eq5}) leads to the unpolarized
momentum distribution, that amounts to keeping only the $I=0$ term.
The terms with $I \neq 0$ appear only when the beam is polarized,
and {\it only} $\ell_x \neq 0$ waves contribute.

We now define the asymmetry 
\begin{equation}
A_i^{M M^\prime} = 
\frac{{\cal Q}_i^M - {\cal Q}_i^{M^\prime}}
     {\sum_M {\cal Q}_i^M}, \hspace{0.7cm} i=1,2
\label{eq6}
\end{equation}
that vanishes when the beam is not polarized.
The deviations from zero are then produced by the $I\neq 0$ terms in
${\cal Q}_i^M$, or, in other words, by the $\ell_x \neq 0$ waves contained 
in the neutron-core subsystem. 

Finally, together with a polarized beam, we consider the possibility of 
measuring the polarization of one of the fragments in the final state.
Then the momentum distribution is similar to eq.(\ref{eq5}), but
containing additional geometrical factors involving the $z$-projection
$\sigma$ of the spin of the polarized particle in the final state (the
expressions are rather extended, and the details will be published in a
coming paper). From the one and two-dimensional momentum distributions 
${\cal Q}_1^{M \sigma}(k_x^\parallel)$ and ${\cal Q}_2^{M \sigma}(k_x^\perp)$ 
we then introduce the asymmetry
\begin{equation}
A_i^{M M^\prime;\sigma \sigma^\prime}=
\frac{ {\cal Q}_i^{M \sigma} 
           - {\cal Q}_i^{M^\prime \sigma^\prime} }
     { \sum_{M \sigma} {\cal Q}_i^{M \sigma} },  \hspace{0.7cm} i=1,2
\label{eq7}
\end{equation} 
where again the deviations from zero are produced by the polarizations.
However, in this case also $s$-waves contribute.

\paragraph*{Results.} We apply the method to the
three-body halo $^{11}$Li ($^9$Li+n+n) fragmentation, where one of the
neutrons is suddenly captured by the target. The two-body potentials used
in the calculations are given in ref.\cite{dima2}, where the
nucleon-nucleon potential is fitted to low energy $s$- and $p$-wave
nucleon-nucleon scattering data, and the neutron-core potential is
adjusted to reproduce the experimental binding energy and mean square 
radius of the three-body system. In all the calculations the lowest 
virtual $s$ state in $^{10}$Li is at 50 keV, and the
lowest $p$-resonance is at 0.5 MeV, according to the experimental data
\cite{zinser}. The rest of the resonances occurs at different energies,
modifying the $p$-content in the two-body subsystem and keeping the right
three-body binding energy.

\begin{figure}[t]
\vspace{1cm}
\epsfxsize=12cm
\epsfysize=7cm
\epsfbox[-100 300 450 650]{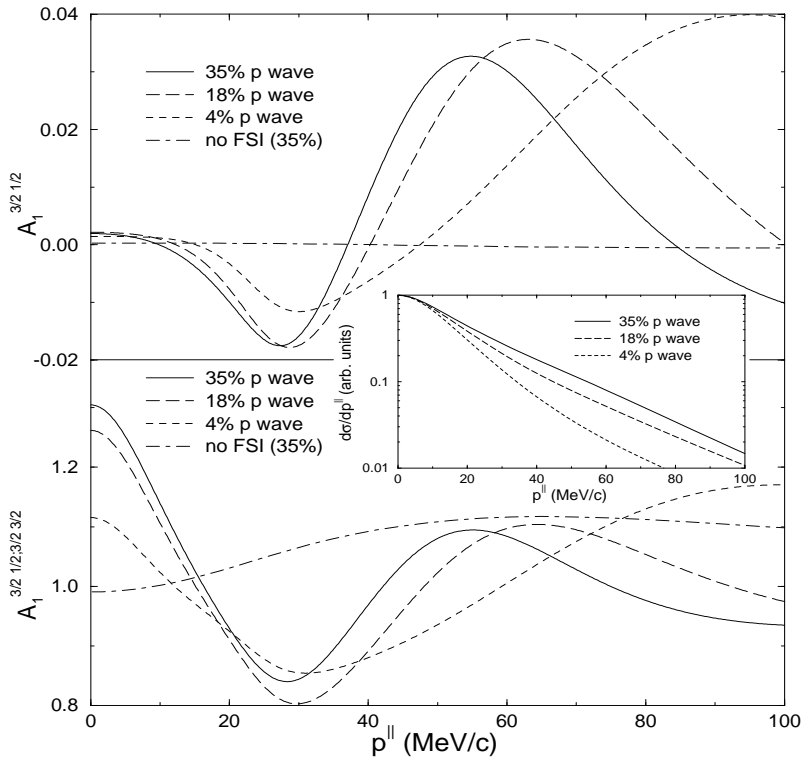}
\vspace{2.2cm}
\caption{\protect\small
Upper part: Asymmetry
$A_1^{\protect\frac{3}{2} \protect\frac{1}{2}}$
for neutron momentum distributions from $^{11}$Li fragmentation reaction
with polarized beam. Three different contents of $p$-wave in the
neutron-core system are considered: 35\% (solid), 18\% (long dashed),
and 4\% (dashed). The  curve for 35\% of $p$-wave without final state
interactions (dot-dashed) is also shown. Lower part: The same as in the
upper part for the asymmetry
$A_1^{\frac{3}{2} \frac{1}{2};\frac{3}{2} \frac{3}{2}}$.
Inner part: Unpolarized one-dimensional neutron momentum distribution
from neutron removal process in $^{11}$Li fragmentation. The
neutron momentum $p$ is relative to the center of mass of the three-body
system.
}
\end{figure}

Let us first focus on the asymmetry $A_1^{\frac{3}{2} \frac{1}{2}}$ for
momentum distributions of the surviving neutron. This asymmetry is shown 
in the upper part of fig.1, where three curves corresponding to different 
$p$-wave content in the neutron-$^9$Li channel (4, 18, and 35\%) are shown. 

From this part of the figure we can conclude the following: 

(i) The deviations from zero of the asymmetry are produced by the $p$-waves 
in the neutron-$^9$Li subsystem. A significant dependence on the $p$-wave 
content of the two-body wave function is observed. Since the asymmetry
is zero when no $p$-waves are present, one could expect smaller 
oscillations  for small $p$-wave content. However, $p$-waves are also
contained in the unpolarized momentum distribution, in such a way that
the less the $p$-wave content, the narrower the momentum distribution (see
the inner part of fig.1). Therefore the denominator of the asymmetry tends
to increase the amplitude of the oscillations for low $p$-wave content
in the two-body wave function. It is then possible, as in
fig.1, to have oscillations with similar amplitude for different values
of the $p$-wave content. 

(ii) If the two-body system has a resonance at a given energy 
$E_{\mbox{\scriptsize res}}$, then the enhanced overlaps in eq.(\ref{eq4}) 
produce a bump in the momentum distribution at 
$k=\sqrt{2 \mu E_{\mbox{\scriptsize res}}}$  ($\mu$ is the 
reduced mass of the two particles). For the same reason the
difference between ${\cal Q}_i^M$ and the unpolarized momentum
distribution is expected to be larger in the regions close
to the resonances in the two-body system. Therefore the asymmetry 
should present larger oscillations for the momentum values corresponding 
to $p$-resonances in $^{10}$Li. In particular, the $p$-resonance 
at 0.5 MeV corresponds to a momentum of 30 MeV/c, that matches with the
position of the 
first peak in the asymmetries. The second $p$-resonance changes
its position from case to case, being placed at 1.5 MeV ($k \sim 50$
MeV/c) for the solid curve, 2 MeV ($k \sim 60$ MeV/c) for the long-dashed
curve, and 6 MeV ($k \sim 100$ MeV/c) for the dashed line. From the
figure it is then clear that the asymmetry gives an indication about the 
level structure of $^{10}$Li. 

(iii) If we consider the three particles in the initial state coupled 
to total orbital angular momentum $L=0$, the polarization of the beam 
affects only the spin part of the wave function. If then none of the 
final spin projections are measured, neglecting the interaction between 
$^9$Li and the remaining neutron (no FSI) leads to momentum distributions 
that are equal to the unpolarized ones, and the asymmetry therefore vanishes
(this can be seen analytically by inserting eqs.(\ref{expan}) and
(\ref{eq2}) into eq.(\ref{eq4}) for $L=0$). In our case, the $^{11}$Li wave 
function contains more than 95\% of $L=0$ configuration, and therefore
the asymmetry neglecting FSI is almost zero (see the dot-dashed line in
the figure). We can therefore conclude that FSI is necessary for the correct
description of the asymmetry.   

\begin{figure}[t]
\vspace{1cm}
\epsfxsize=12cm
\epsfysize=7cm
\epsfbox[-100 300 450 650]{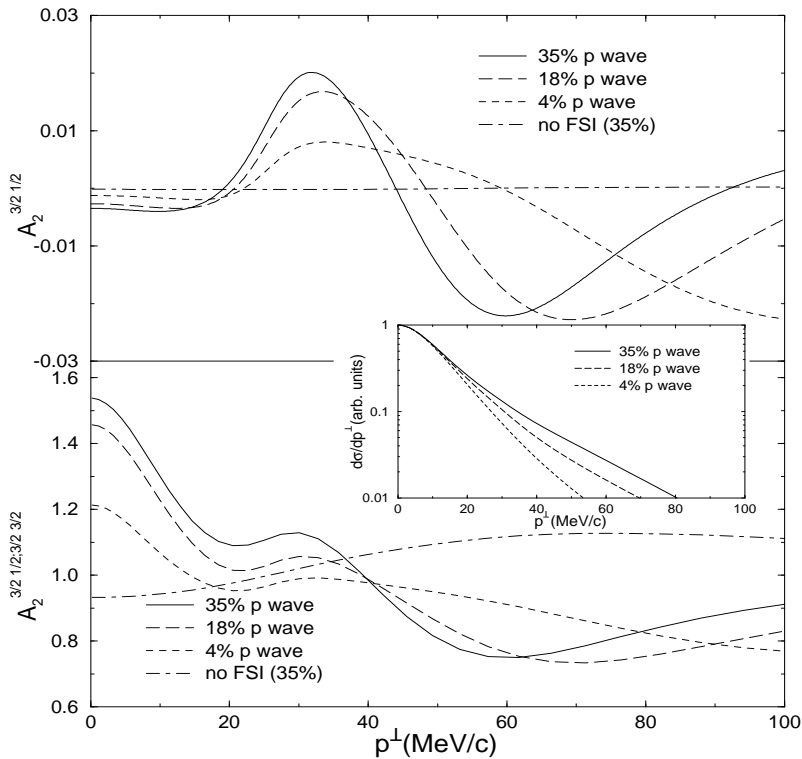}
\vspace{2.2cm}
\caption{\protect\small
The same as in fig.1 for two-dimensional neutron momentum
distributions.
}
\end{figure}

We turn now to the case where the polarization of the surviving core also
is measured. We assume that the core is longitudinally polarized and 
consider the asymmetry 
$A_1^{\frac{3}{2} \frac{1}{2};\frac{3}{2} \frac{3}{2}}$.
In the lower part of fig.1 we plot this asymmetry for the same three
cases we considered above. The three points we already emphasized 
are valid again. First a clear dependence on the $p$-wave content is 
observed even though the $s$-waves also appear in the
numerator of the asymmetry. The effect of $s$-waves is however
confined to the low momentum region, and for momenta larger than 20
MeV/c only $p$-waves contribute. Therefore the minima and maxima of the 
asymmetry again match with the $p$-resonances in $^{10}$Li. Finally 
the dot-dashed line shows that FSI is also
essential in this case. However as the polarization of the core is now 
measured, even $L=0$ terms in the three-body wave function contribute to the
asymmetry. This asymmetry is then an observable with similar characteristics 
as $A_1^{\frac{3}{2} \frac{1}{2}}$, but 
with the advantage that the oscillations are larger, and therefore easier 
to observe experimentally.

In fig.2 we show the same calculations as in fig.1, but for 
two-dimensional neutron momentum distributions. The curves show features
similar to the curves in fig.1. Our conclusions about the behaviour of
the asymmetries are therefore rather general and not restricted to only 
one kind of momentum distributions.

\paragraph*{Summary and conclusions.} The aim of this letter is to
investigate polarizations in three-body halo fragmentation reactions at
high energies and light targets. We use the sudden approximation where
the initial state is a genuine three-body wave function and the final
state is a distorted wave in the same potential between the two
surviving particles as used in the
three-body calculation. We have defined several 
asymmetries from which the polarization effects can be analyzed.    

We applied the method to $^{11}$Li fragmentation, and computed one and
two-dimensional neutron momentum distributions. We have found that the
polarization observables are more sensitive to the $p$-wave content 
in the neutron-core subsystem than unpolarized cross sections. Furthermore
they provide a unique information of the level structure of the two-body
subsystem. 
We also observed that FSI
is the main cause for the behaviour of the asymmetries, and FSI must
necessarily be included in the calculation. These features have been
found for fragmentation reactions with polarized beam, and also when 
simultaneously the polarization of the core is measured. The advantage
of the second case is that the amplitude of the oscillations observed in
the asymmetries is clearly larger. Finally we have shown that these
conclusions are valid for both one and two-dimensional momentum 
distributions.   
 
{\bf Acknowledgments} One of us (E.G.) acknowledges support from the
European Union through the Human Capital and Mobility program contract
nr. ERBCHBGCT930320.


\begin{thebibliography}{99}
\bibitem{han95} {\sc P.G. Hansen, A.S. Jensen}, and {\sc B. Jonson}, 
{\it Ann. Rev. Nucl. Part. Sci.}, {\bf 45} (1995) 591.
\bibitem{dima1} {\sc D.V. Fedorov, A.S. Jensen}, and {\sc K. Riisager}, 
{\it Phys. Rev. C}, {\bf 49} (1994) 201. 
\bibitem{zhukov} {\sc M.V. Zhukov, B.V. Danilin, D.V. Fedorov, J.M. Bang, I.J.
Thompson}, and {\sc J.S. Vaagen}, 
{\it Phys. Rep.}, {\bf 231} (1993) 151. 
\bibitem{anne} {\sc R. Anne} {\it et al.}, 
{\it Phys. Lett. B}, {\bf 250} (1990) 19.
\bibitem{zinser} {\sc M. Zinser} {\it et al.}, 
{\it Phys. Rev. Lett.}, {\bf 75} (1995) 1719.
\bibitem{nilsson} {\sc T. Nilsson} {\it et al.}, 
{\it Europhys. Lett.}, {\bf 30} (1995) 19.
\bibitem{humbert} {\sc F. Humbert} {\it et al.}, 
{\it Phys. Lett. B}, {\bf 347} (1995) 198.
\bibitem{thompson} {\sc I.J. Thompson} and {\sc M.V. Zhukov}, 
{\it Phys. Rev. C}, {\bf 49} (1994) 1904. 
\bibitem{korshe} {\sc A.A. Korsheninnikov} and {\sc T. Kobayashi}, 
{\it Nucl. Phys. A}, {\bf 567} (1994) 97.
\bibitem{edu} {\sc E. Garrido, D.V. Fedorov}, and {\sc A.S. Jensen}, 
in press in {\it Phys. Rev. C}, {\bf 53} (1996),
http://xxx.lanl.gov/abs/nucl-th/9603028
\bibitem{serber} {\sc R. Serber}, {\it Phys. Rev.}, {\bf 72} (1947) 1008. 
\bibitem{satchler} {\sc G.R. Satchler}, {\it Direct Nuclear Reactions, 
(Oxford University Press, Oxford, 1983)}. 
\bibitem{dima2} {\sc D.V. Fedorov, E. Garrido}, and {\sc A.S. Jensen}, 
{\it Phys. Rev. C}, {\bf 51} (1995) 3052.

\end{thebibliography}
\end{document}